\documentclass[aip,reprint,amsmath,amssymb]{revtex4-1}

\usepackage{graphicx}

\begin{document}

\title{High Quality Factor Platinum Silicide Microwave Kinetic Inductance Detectors} 

\author{P. Szypryt*}
\email[]{pszypryt@physics.ucsb.edu}
\affiliation{Department of Physics, University of California, Santa Barbara, California 93106, USA}

\author{B. A. Mazin}
\affiliation{Department of Physics, University of California, Santa Barbara, California 93106, USA}

\author{G. Ulbricht}
\affiliation{Department of Physics, University of California, Santa Barbara, California 93106, USA}

\author{B. Bumble}
\affiliation{NASA Jet Propulsion Laboratory, Pasadena, California 91109, USA}


\author{S. R. Meeker}
\affiliation{Department of Physics, University of California, Santa Barbara, California 93106, USA}

\author{C. Bockstiegel}
\affiliation{Department of Physics, University of California, Santa Barbara, California 93106, USA}

\author{A. B. Walter}
\affiliation{Department of Physics, University of California, Santa Barbara, California 93106, USA}

\date{\today}

\begin{abstract}
We report on the development of Microwave Kinetic Inductance Detectors (MKIDs) using platinum silicide as the sensor material.  MKIDs are an emerging superconducting detector technology, capable of measuring the arrival times of single photons to better than two microseconds and their energies to around ten percent.  Previously, MKIDs have been fabricated using either sub-stoichiometric titanium nitride or aluminum, but TiN suffers from spatial inhomogeneities in the superconducting critical temperature and Al has a low kinetic inductance fraction, causing low detector sensitivity.  To address these issues, we have instead fabricated PtSi microresonators with superconducting critical temperatures of 944$\pm$12~mK and high internal quality factors ($Q_i \gtrsim 10^6$).  These devices show typical quasiparticle lifetimes of $\tau_{qp} \approx 30$--$40~\mu$s and spectral resolution, $R = \lambda / \Delta \lambda$, of 8 at 406.6~nm. We compare PtSi MKIDs to those fabricated with TiN and detail the substantial advantages that PtSi MKIDs have to offer.
\end{abstract}

\pacs{}

\maketitle 

Microwave Kinetic Inductance Detectors (MKIDs\cite{Day2003}) are low-temperature detectors capable of measuring the arrival times of single photons to better than two microseconds and their energies to around ten percent. MKID operation depends on the kinetic inductance effect\cite{Mattis1958}, an additional inductance term which can be exploited for single photon detection. Cooper Pairs are broken when a superconductor below its $T_c$ absorbs a photon, creating a population of unpaired electrons called quasiparticles.  The sudden decrease in Cooper Pair density temporarily increases the kinetic inductance of the superconducting film.  If a thin film superconductor is lithographically patterned into a microresonator, a photon absorption event will then act to momentarily decrease the resonant frequency of the microresonator.  The energy of the incident photon is proportional to the number of broken Cooper Pairs, and therefore the change in frequency, giving MKIDs spectral resolution. MKIDs are naturally multiplexed by assigning each microresonator in an array a unique frequency during lithography. They can then be read out\cite{McHugh2012} using a frequency domain multiplexing scheme. With this method employed on the latest generation of digital microwave electronics, thousands detectors can be coupled to and read out using a single microwave transmission line\cite{Strader2016b}.

Because the bandgap $\Delta$ of a superconductor is roughly $10^4$ times smaller than that of the silicon used in conventional charge-coupled devices (CCDs), MKIDs are capable of detecting the long wavelength photons that would typically pass right through a CCD.  MKIDs can be operated over a broad wavelength range and a handful of MKID astronomy instruments in the submillimeter\cite{Schlaerth2012,Calvo2016} and the ultraviolet, optical, and near-infrared (UVOIR)\cite{Mazin2013,Meeker2015} wavelength bands have been commissioned. Although our work is primarily focused on the UVOIR regime, MKIDs operating at all wavelengths benefit from advances in the superconducting sensor layer.

We have chosen platinum silicide as the superconductor in our devices in order to avoid the problems observed in previously fabricated sub-stoichiometric titanium nitride\cite{Leduc2010,Vissers2013} and aluminum\cite{Goupy2016} MKIDs. In TiN MKIDs, thin films are deposited by reactively sputtering off of a Ti target in a nitrogen atmosphere, and the N$_2$ flow rate can be used to control the TiN stoichiometry and $T_c$. Unfortunately, for a given Ti sputtering power, the $T_c$ is highly sensitive to the N to Ti ratio near the desired $T_c$ of 1~K\cite{Leduc2010}.  This particular method leads to local and long-range variations in $T_c$ (and sheet inductance) across the wafer\cite{Vissers2013}, causing the microresonators to shift away from their design frequencies.  Microresonators overlapping in frequency are unable to be read out, lowering the overall detector yield. This issue is only exacerbated in large format arrays where microresonators need to be spaced closely together in frequency space to stay within the bandwidth of the readout electronics.  Attempts have been made to make TiN more uniform through multilayer stacking of TiN/Ti/TiN\cite{Vissers2013}, but this technique is incompatible with several essential UVOIR MKID fabrication steps.  Al, on the other hand, can be deposited very uniformly but has a short London penetration depth (50 nm, compared to ${\sim}1~\mu$m for TiN), indicating it has a high Cooper pair density, therefore explaining its low kinetic inductance fraction.  In order to maintain a high sensitivity (a large phase shift per number of broken Cooper Pairs), the Al film needs to be very thin ($\lesssim$~5~nm), but this creates severe issues with oxidation and photon absorption.  On top of this, Al is very reflective to UVOIR photons, making overall detector quantum efficiency even lower. Al is also known to exhibit long quasiparticle recombination times of $\gtrsim$~1~ms after photon hits, making the material unsuitable for applications that require single photon detection at high count rates ($\gtrsim100$~counts/s).

PtSi was chosen as a candidate material for the MKID superconductor for a number of reasons. Deposition of PtSi is fairly well understood, and its thin film, room temperature electronic and optical properties have been measured\cite{Bentmann2008}. Many of the PtSi superconducting properties have also been studied\cite{Oto1994}, including a measured $T_c$ value of $\sim$1~K which decreases with thickness for films thinner than 50~nm.  Typically, PtSi films are formed by first depositing Pt on a high resistivity Si substrate.  The sample is then annealed at temperatures above $300^\circ$~C, causing the Pt to diffuse into the Si and form a PtSi film of roughly twice the initial Pt thickness.  With this method, the thermally stable state consists of one Pt atom for every Si atom, and fortuitously this happens to be a state with a $T_c$ of $\sim$1~K for films thicker than 50~nm\cite{Oto1994}.  This method was employed to fabricate the initial PtSi MKIDs, but the resulting resonators had low internal quality factors, $Q_i$ due to excess Pt diffusion deep into the Si substrate\cite{Szypryt2015}.  A different method was designed to improve $Q_i$ by utilizing a sapphire substrate as a Pt diffusion barrier. The remainder of this discussion will involve PtSi MKIDs grown on sapphire substrates.

To begin, a sputter system with a base pressure of $\sim$$10^{-7}$ Torr was used to deposit a 30~nm Pt film on a one-side polished C-plane (0001) sapphire wafer. A 45~nm Si film was then sputter deposited on top of the Pt film and the sample was annealed at $500^\circ$~C for 25~minutes, resulting in a PtSi film of roughly 60~nm.  These steps were all done in situ without breaking vacuum. The resulting PtSi film was then patterned with simple one layer MKID test structures using a deep UV stepper. The film was dry-etched using an inductively coupled plasma etcher with a combination of Ar, Cl$_2$, and CF$_4$ gasses. The resonators were designed with 40$\times$40~$\mu$m meandered inductors and 200$\times$160~$\mu$m interdigitated capacitors with variable leg lengths, putting the resonators into a 4--8 GHz frequency range.

Low temperature testing to measure the PtSi superconducting properties was done using a dilution refrigerator controlled at 100~mK.  A $T_c$ of 944$\pm$12~mK was measured using a DC resistance testbed during the cool down.  At 100~mK, a vector network analyzer was used to find the locations of the resonators in frequency space.  By comparing the locations of these resonators to their design frequencies, and noting a design sheet inductance of 10.0~pH/$\square$, we measured the average local PtSi sheet inductance to be 8.2$\pm$0.2~pH/$\square$.

Once the exact locations of the resonators were known, an analog microwave readout system was used to send and receive probe tones scanning the frequency space around the device resonant frequencies. Resonators were swept between $-107$ and $-119$~dBm of power, and for all subsequent low temperature measurements, individual resonators were operated at the highest possible power before the onset of non-linearity. The complex transmission was measured near resonance and individual resonators were fit using a capacitively coupled LC resonator model\cite{Gao2008b}. Of the resonators with a good fit, a mean $Q_i$ value of $1.06\times10^6$ was calculated.  Because the total quality factor is dominated by the smallest of $Q_i$ and $Q_c$ (coupling quality factor), there is a large inherent spread in the fitted $Q_i$ when it is much larger than $Q_c$ (designed to be between 40,000--50,000).  In this case, a standard deviation in $Q_i$ of $5.5\times10^5$ was measured. The mean measured $Q_c$ for these resonators was 38,000. These values of $Q_i$ are among the highest seen in superconducting microresonators among a wide variety of materials.

\begin{figure}[h]
	\centering
	\includegraphics[width=\linewidth]{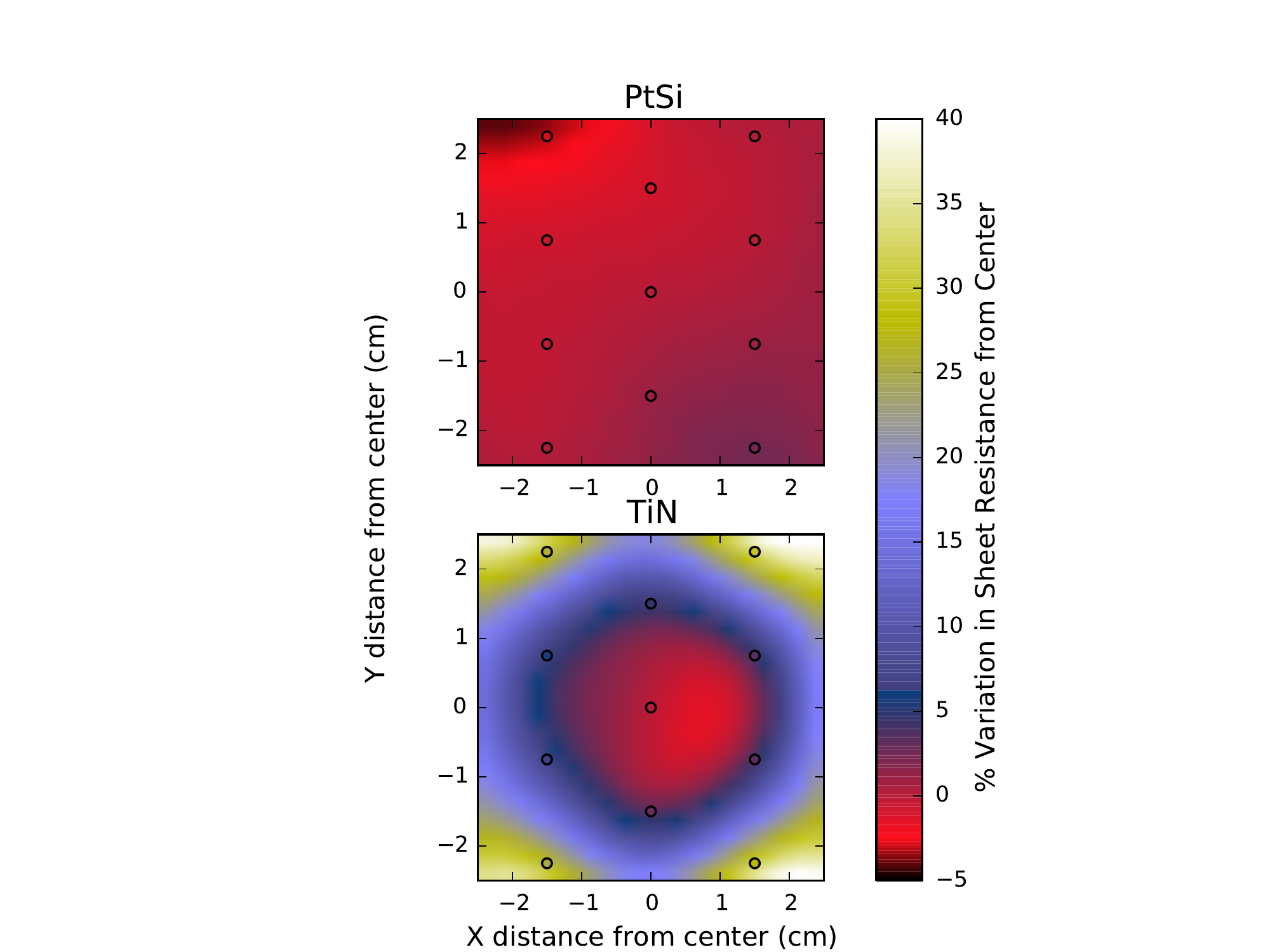}
	\caption{Percent variation in sheet resistance of PtSi and TiN thin films from the center of a 4" wafer. Measurements were done at the locations of the circles and the filled contour map was generated using a radial basis function interpolation.}
	\label{fig:ResMap}
\end{figure}

\begin{figure}[h]
	\centering
	\includegraphics[width=\linewidth]{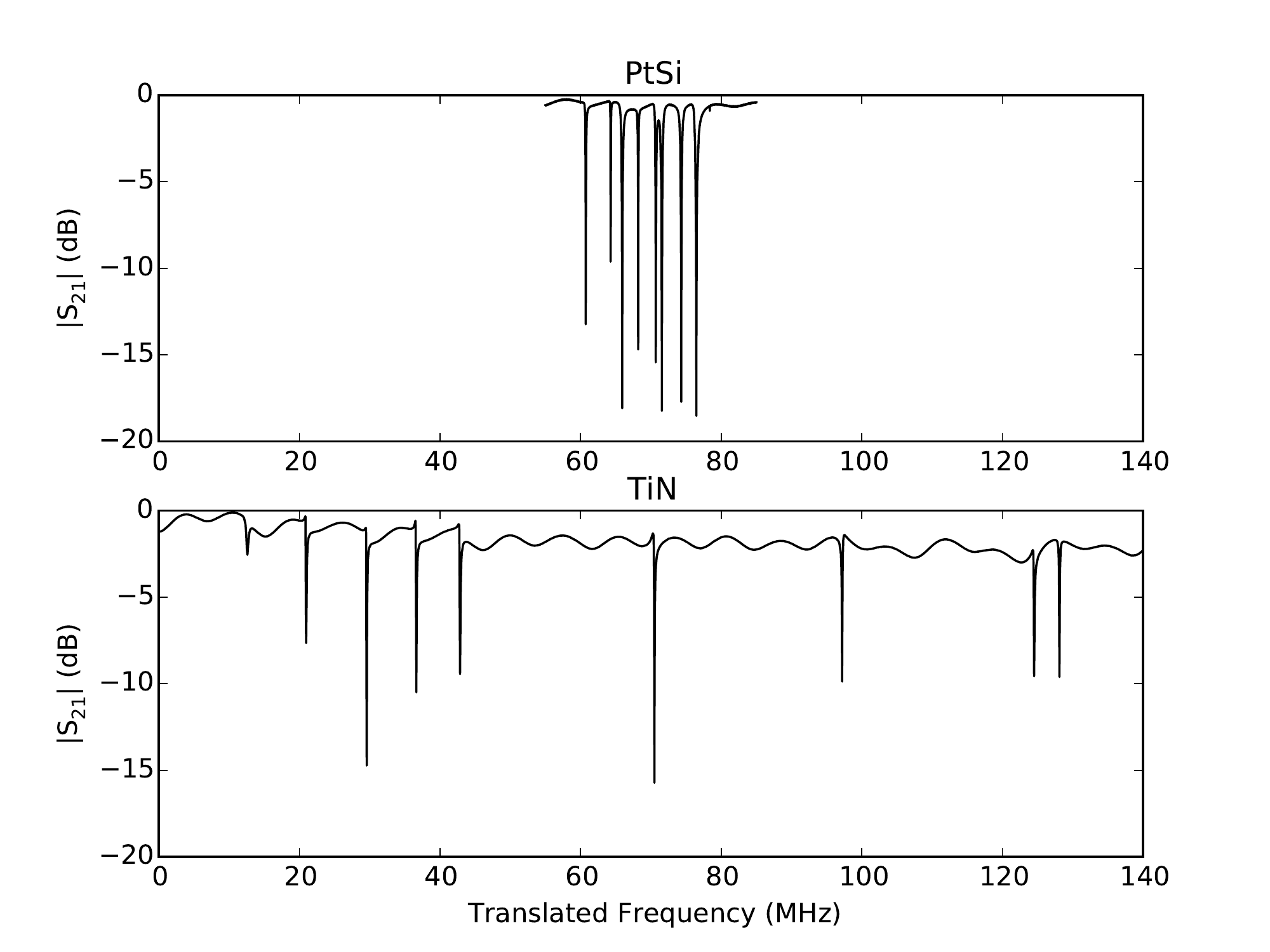}
	\caption{Wide frequency sweep of identical resonator structures designed at 2~MHz spacing for PtSi (top) and TiN (bottom). The base line has been translated to 0~dB for clarity.  The frequency has also been shifted to 0~MHz for an easier comparison of the TiN and PtSi devices.  An offset of 4.845~GHz is present for PtSi and 3.620~GHz for TiN. The seventh PtSi resonator in the sequence is missing due to a photomask error.}
	\label{fig:ResonatorSpacing}
\end{figure}

The uniformity of the PtSi devices was also measured and compared to that of devices fabricated using sub-stoichiometric TiN, as non-uniformity in $T_c$ is the most significant problem currently plaguing TiN MKIDs. For these superconducting films, the uniformity in room temperature sheet resistance can often be used as a rough proxy for uniformity in $T_c$. Figure~\ref{fig:ResMap} shows the percent variation of film sheet resistance from the center of a 4" wafer for PtSi and TiN.  These measurements show almost an order of magnitude better uniformity in PtSi than in TiN.  The fabrication mask also contains a group of resonators closely spaced in frequency with a designed spacing of 2~MHz.  This is meant to resemble the 2~MHz spacing used between detectors in large format MKID arrays. Complex transmission magnitudes for identical structures made with PtSi and TiN are shown in Figure~\ref{fig:ResonatorSpacing}.  Note that the PtSi resonators spread out to around 20~MHz of bandwidth, whereas the TiN resonators spread to 120~MHz.  The total bandwidth of the 9 resonators was designed to be 16~MHz. It can be seen that the frequency variations are less pronounced in PtSi than in TiN, allowing for finer multiplexing and a smaller number of frequency collisions.

\begin{figure}[h]
	\centering
	\includegraphics[width=\linewidth]{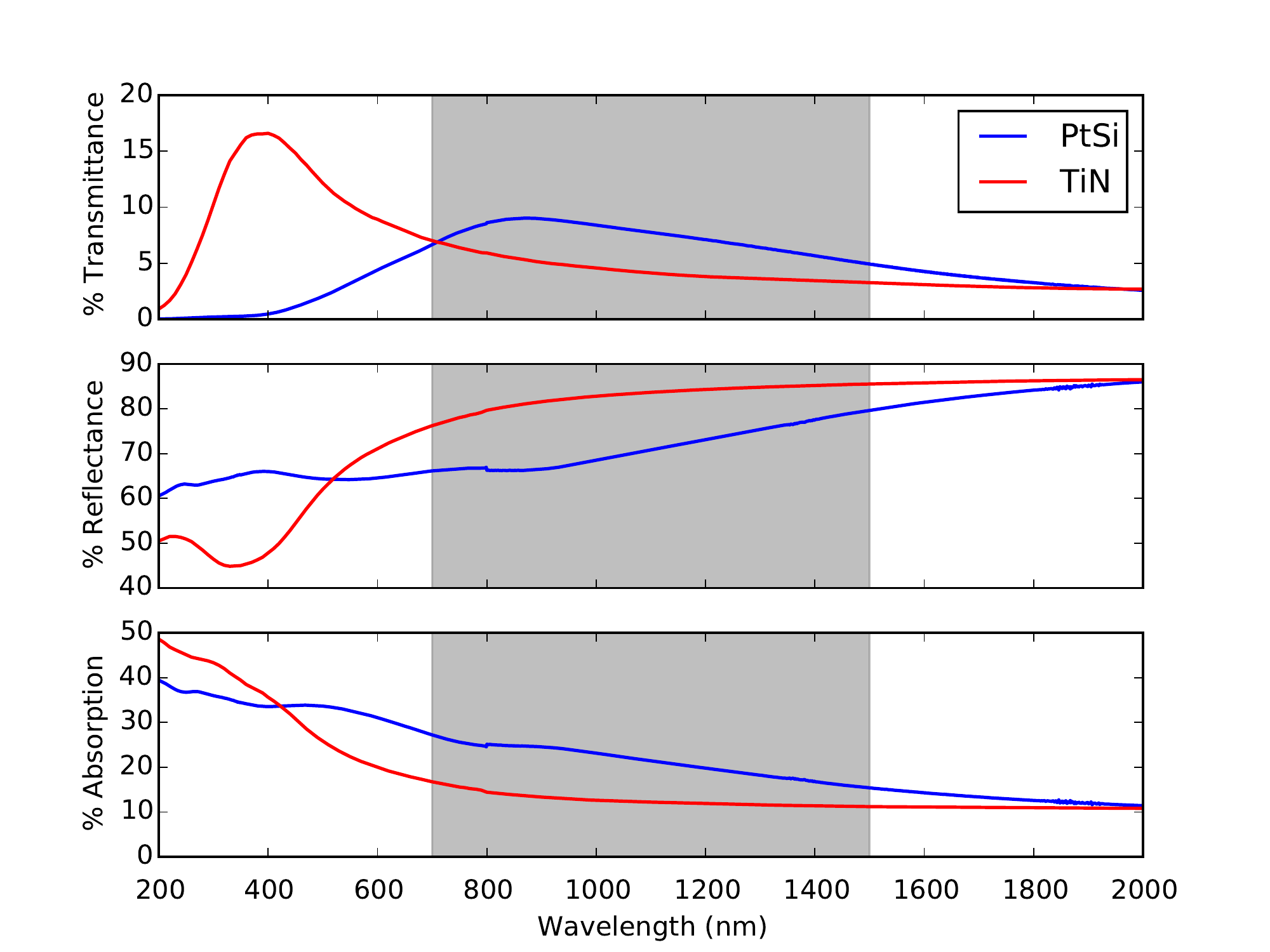}
	\caption{Optical transmittance, reflectance, and absorption measurements of unpatterned 60~nm PtSi (blue) and sub-stoichiometric 1~K~T$_C$ TiN (red) thin films on a sapphire substrates. The shaded region represents the wavelength band of the upcoming MKID instrument, DARKNESS\cite{Meeker2015}. The slight discontinuity at 800~nm is the result of the spectrophotometer switching its light source.}
	\label{fig:QE}
\end{figure}

Room temperature transmittance, reflectance, and absorption measurements of unpatterned PtSi films were done using an Agilent Cary 5000 wideband spectrophotometer and are shown in Figure~\ref{fig:QE}.  This data is used to determine an upper limit on quantum efficiency of our detectors over the wide wavelength range in which they operate.  The data was compared to that of a sub-stoichiometric TiN film typically used in MKIDs. TiN has better absorption than PtSi at shorter wavelengths, however, at wavelengths over 425~nm, PtSi starts to outperform the TiN.  This will be useful as future UVOIR MKID exoplanet imaging instruments\cite{Meeker2015} start pushing further into the near-IR regime.

\begin{figure}[h]
	\centering
	\includegraphics[width=\linewidth]{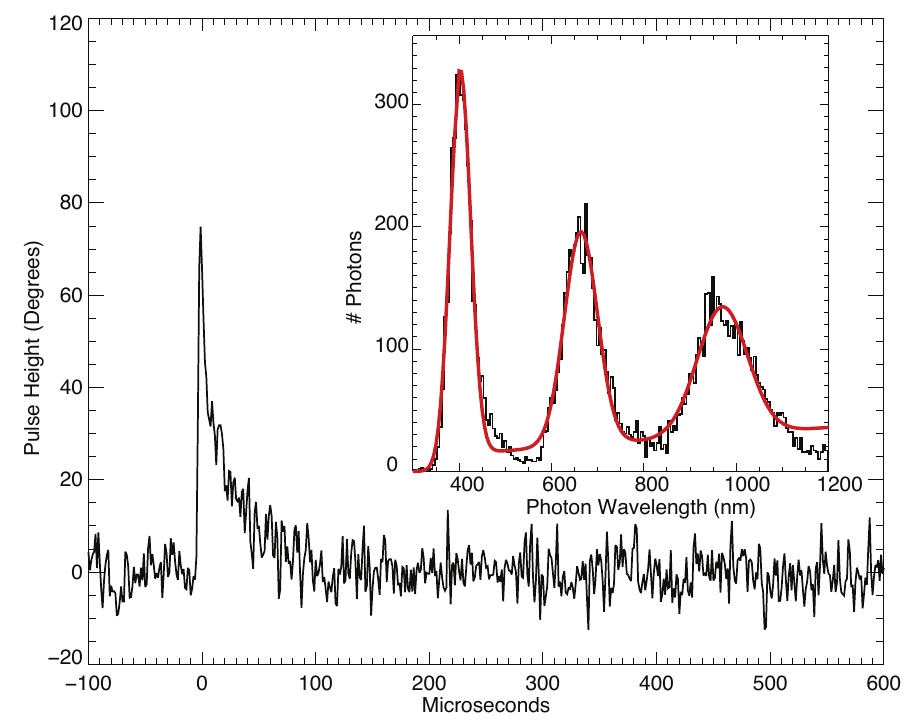}
	\caption{A typical single 671.0 nm photon being absorbed by a PtSi MKID with f$_0$=4.876 GHz, Q$_c$=15,700, and Q$_i$=147,300.  This particular resonator was probed with a power level of $-108$~dBm. The fitted quasiparticle recombination time is $36\pm2$~$\mu$s. Inset: The spectrum of the same MKID that has been illuminated with 406.6, 671.0, and 982.1 nm lasers. The data is transformed from phase height into wavelength using these known laser wavelengths\cite{vanEyken2015}.  The red line is a fit with three Gaussians and a linear background term, yielding a nearly uniform spectral resolution R=$\lambda/\Delta\lambda$=8 across the entire 400--1000 nm range.}
	\label{fig:PulseHeightHistogram}
\end{figure}

Photon testing was done in an adiabatic demagnetization refrigerator with optical access, also controlled at 100~mK.  As explained earlier, when a photon hits a superconducting microresonator, it shifts the resonant frequency of the microresonator.  One can determine the energy of the photon by measuring this frequency shift, but in the MKID readout, each resonator is read out using only a single, unique probe tone, making this type of measurement difficult.  The frequency shift will, however, cause a change in the amplitude and phase of the complex transmission at the particular probe tone. Measuring the phase shift tends to result in higher signal-to-noise ratios, and the phase can also be converted to energy in a straightforward way\cite{vanEyken2015}.  In order to measure the spectral resolution of the device it was illuminated with three lasers of known wavelength and a histogram was created, as shown in Figure~\ref{fig:PulseHeightHistogram}. The histogram was fit to a model of three Gaussians (one for each laser) and a linear background term.  Typical PtSi resonators had spectral resolutions R=$\lambda/\Delta\lambda$=8 at 406.6 nm and quasiparticle recombination times $\tau_{qp} \approx 30$--$40~\mu$s. The spectral resolution in the PtSi device was very similar to the results achieved in our best TiN resonators of similar geometry.  At these levels, the spectral resolution is being limited by a combination of amplifier and two-level system (TLS\cite{Gao2008}) noise rather than the properties of the superconducting material.

\begin{figure}[h]
	\centering
	\includegraphics[width=\linewidth]{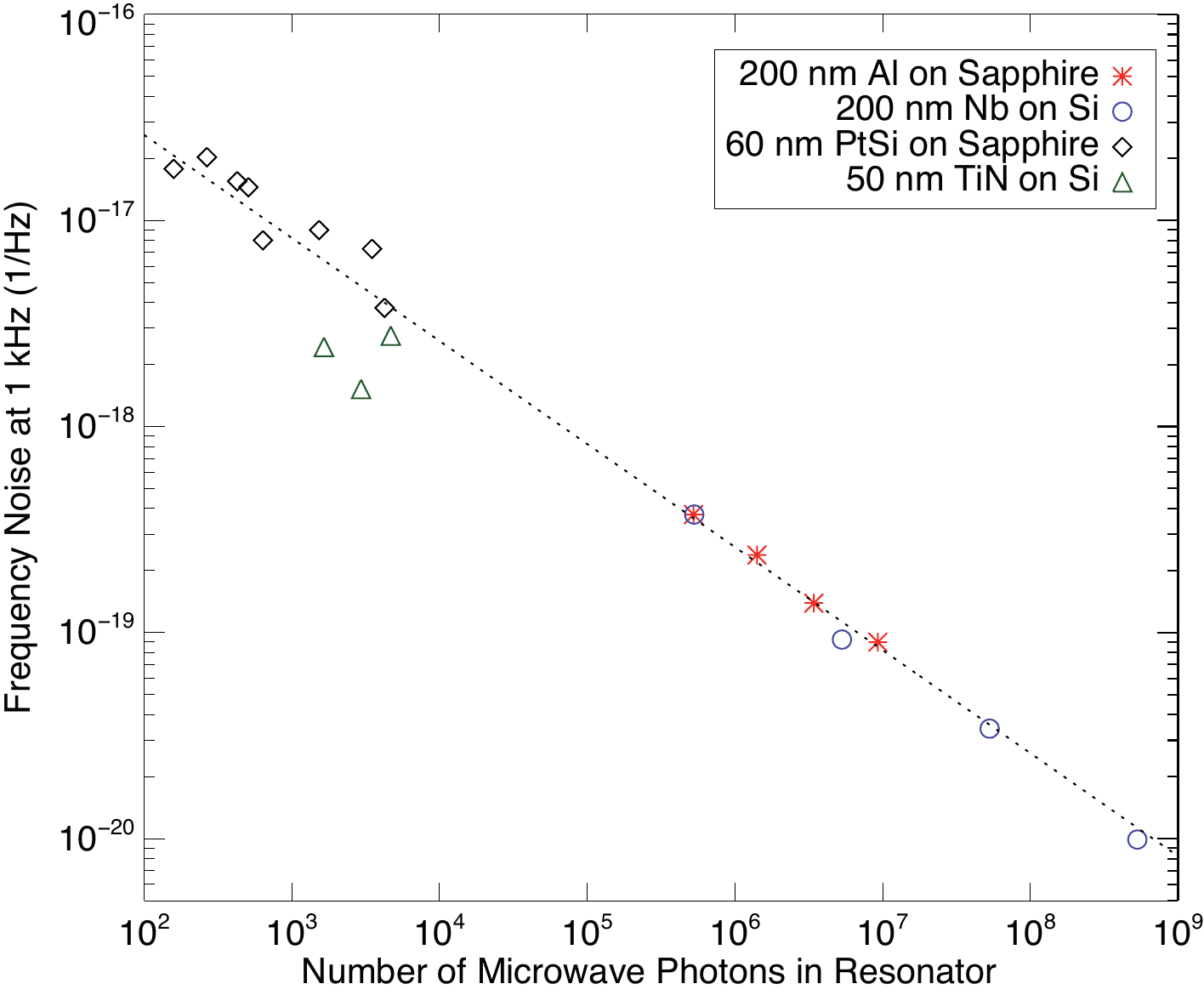}
	\caption{Fractional frequency noise (1/Hz) at 1 kHz as a function of the number of photons in a $\lambda/4$ CPW resonator. PtSi and TiN resonator measurements are plotted together with archival data of measurements of Nb on Si and Al on sapphire resonators\cite{Noroozian2012}.  The dashed line has a power law slope of -0.5 to guide the eye with the expected change in noise as a function of photon number. The powers in PtSi and TiN are significantly lower than previous measurements because these devices are optimized for optical single photon detection, whereas previous samples were made with thicker films with a lower kinetic inductance fraction, yielding less responsive resonators.}
	\label{fig:NoiseComp}
\end{figure}

Noise measurements, shown in Figure~\ref{fig:NoiseComp}, were made using $\lambda/4$ coplanar waveguide (CPW) resonators with a 3~$\mu$m center strip and 2~$\mu$m gaps.  The data is compared with previous measurements\cite{Noroozian2012} from Al on Sapphire and Nb on Si resonators, and the results match expectations extremely well despite the number of photons in the resonator (a more generic proxy for the resonator readout power) differing by nearly four orders of magnitude.  This indicates that the PtSi is adding no extra TLS noise compared to standard MKID materials.  The graph also includes some data from similar $\lambda/4$ CPW resonators made from TiN on Si, and these resonators do appear to be systematically slightly less noisy than the Al, Nb, and PtSi.  This lower noise may be related to the unusually clean interface between HF-dipped Si and TiN, as TiN on Si lithographed microwave resonators have shown exceptionally high $Q_i$ ($> 10^7$)~\cite{Leduc2010}.

MKIDs for UV, optical, and near-IR photon detection are extremely promising detectors for astronomical observations, but they need improvements in three distinct areas: pixel yield, spectral resolution and quantum efficiency. First, the much more uniform nature of PtSi, shown in Figures \ref{fig:ResMap} and \ref{fig:ResonatorSpacing}, indicates that it exhibits little frequency scatter, which should dramatically improve pixel yield. Next, we see a significant discrepancy between the observed R and the R predicted from the formalism of optimal filtering~\cite{Golwala2000} in the best TiN on Si MKIDs.  We believe this is due to the non-uniformity of the TiN bandgap extending down to very small spatial scales\cite{Sacepe2008} and the short quasiparticle diffusion lengths causing a different number of quasiparticles to be generated at different locations along the length of the inductor. In PtSi on sapphire we have seen a good match between predicted and observed R.  This means as we improve our detector and readout design to reduce TLS and HEMT noise we expect to see improvements in R beyond the hard limit of R$\approx$10 we see in TiN. Finally, PtSi offers significantly higher quantum efficiency than TiN in the critical near-IR area where MKIDs will be used for exoplanet direct imaging work behind adaptive optics systems\cite{Meeker2015}, as shown in Figure~\ref{fig:QE}.

PtSi on sapphire appears to be a significant breakthrough for photon counting MKIDs, and development of 10 kpix PtSi on sapphire MKID arrays is proceeding. There are still paths forward for further raising the $Q_i$, such as more advanced sapphire substrate cleaning procedures. In addition, going to ultra-high vacuum sputter chambers with lower base pressures will help to reduce impurities in the PtSi films, further improving $Q_i$.\\

This work was supported by a NASA Space Technology Research Fellowship (NSTRF). Fabrication was done in the UCSB Nanofabrication Facility.  The authors would like to thank the Las Cumbres Observatory Global Telescope (LCOGT) network for assisting in broadband quantum efficiency measurements and Omid Noroozian for providing archival noise data.

\bibliographystyle{apsrev}

\begin{thebibliography}{20}
\expandafter\ifx\csname natexlab\endcsname\relax\def\natexlab#1{#1}\fi
\expandafter\ifx\csname bibnamefont\endcsname\relax
  \def\bibnamefont#1{#1}\fi
\expandafter\ifx\csname bibfnamefont\endcsname\relax
  \def\bibfnamefont#1{#1}\fi
\expandafter\ifx\csname citenamefont\endcsname\relax
  \def\citenamefont#1{#1}\fi
\expandafter\ifx\csname url\endcsname\relax
  \def\url#1{\texttt{#1}}\fi
\expandafter\ifx\csname urlprefix\endcsname\relax\def\urlprefix{URL }\fi
\providecommand{\bibinfo}[2]{#2}
\providecommand{\eprint}[2][]{\url{#2}}

\bibitem[{\citenamefont{Day et~al.}(2003)\citenamefont{Day, Leduc, Mazin,
  Vayonakis, and Zmuidzinas}}]{Day2003}
\bibinfo{author}{\bibfnamefont{P.~K.} \bibnamefont{Day}},
  \bibinfo{author}{\bibfnamefont{H.~G.} \bibnamefont{Leduc}},
  \bibinfo{author}{\bibfnamefont{B.~A.} \bibnamefont{Mazin}},
  \bibinfo{author}{\bibfnamefont{A.}~\bibnamefont{Vayonakis}},
  \bibnamefont{and}
  \bibinfo{author}{\bibfnamefont{J.}~\bibnamefont{Zmuidzinas}},
  \bibinfo{journal}{Nature} \textbf{\bibinfo{volume}{425}},
  \bibinfo{pages}{817} (\bibinfo{year}{2003}).

\bibitem[{\citenamefont{Mattis and Bardeen}(1958)}]{Mattis1958}
\bibinfo{author}{\bibfnamefont{D.~C.} \bibnamefont{Mattis}} \bibnamefont{and}
  \bibinfo{author}{\bibfnamefont{J.}~\bibnamefont{Bardeen}},
  \bibinfo{journal}{Phys. Rev.} \textbf{\bibinfo{volume}{111}},
  \bibinfo{pages}{412} (\bibinfo{year}{1958}).

\bibitem[{\citenamefont{McHugh et~al.}(2012)\citenamefont{McHugh, Mazin,
  Serfass, Meeker, O’Brien, Duan, Raffanti, and Werthimer}}]{McHugh2012}
\bibinfo{author}{\bibfnamefont{S.}~\bibnamefont{McHugh}},
  \bibinfo{author}{\bibfnamefont{B.~A.} \bibnamefont{Mazin}},
  \bibinfo{author}{\bibfnamefont{B.}~\bibnamefont{Serfass}},
  \bibinfo{author}{\bibfnamefont{S.}~\bibnamefont{Meeker}},
  \bibinfo{author}{\bibfnamefont{K.}~\bibnamefont{O’Brien}},
  \bibinfo{author}{\bibfnamefont{R.}~\bibnamefont{Duan}},
  \bibinfo{author}{\bibfnamefont{R.}~\bibnamefont{Raffanti}}, \bibnamefont{and}
  \bibinfo{author}{\bibfnamefont{D.}~\bibnamefont{Werthimer}},
  \bibinfo{journal}{Rev. Sci. Instrum.} \textbf{\bibinfo{volume}{83}},
  \bibinfo{eid}{044702} (\bibinfo{year}{2012}).

\bibitem[{\citenamefont{Strader}(2016)}]{Strader2016b}
\bibinfo{author}{\bibfnamefont{M.~J.} \bibnamefont{Strader}}, Ph.D. thesis,
  \bibinfo{school}{University of California, Santa Barbara}
  (\bibinfo{year}{2016}).

\bibitem[{\citenamefont{Schlaerth et~al.}(2012)\citenamefont{Schlaerth, Czakon,
  Day, Downes, Duan, Glenn, Golwala, Hollister, LeDuc, Maloney
  et~al.}}]{Schlaerth2012}
\bibinfo{author}{\bibfnamefont{J.~A.} \bibnamefont{Schlaerth}},
  \bibinfo{author}{\bibfnamefont{N.~G.} \bibnamefont{Czakon}},
  \bibinfo{author}{\bibfnamefont{P.~K.} \bibnamefont{Day}},
  \bibinfo{author}{\bibfnamefont{T.~P.} \bibnamefont{Downes}},
  \bibinfo{author}{\bibfnamefont{R.}~\bibnamefont{Duan}},
  \bibinfo{author}{\bibfnamefont{J.}~\bibnamefont{Glenn}},
  \bibinfo{author}{\bibfnamefont{S.~R.} \bibnamefont{Golwala}},
  \bibinfo{author}{\bibfnamefont{M.~I.} \bibnamefont{Hollister}},
  \bibinfo{author}{\bibfnamefont{H.~G.} \bibnamefont{LeDuc}},
  \bibinfo{author}{\bibfnamefont{P.~R.} \bibnamefont{Maloney}},
  \bibnamefont{et~al.}, \bibinfo{journal}{J. Low Temp. Phys.}
  \textbf{\bibinfo{volume}{167}}, \bibinfo{pages}{347} (\bibinfo{year}{2012}).

\bibitem[{\citenamefont{Calvo et~al.}(2016)\citenamefont{Calvo, Beno{\^i}t,
  Catalano, Goupy, Monfardini, Ponthieu, Barria, Bres, Grollier, Garde
  et~al.}}]{Calvo2016}
\bibinfo{author}{\bibfnamefont{M.}~\bibnamefont{Calvo}},
  \bibinfo{author}{\bibfnamefont{A.}~\bibnamefont{Beno{\^i}t}},
  \bibinfo{author}{\bibfnamefont{A.}~\bibnamefont{Catalano}},
  \bibinfo{author}{\bibfnamefont{J.}~\bibnamefont{Goupy}},
  \bibinfo{author}{\bibfnamefont{A.}~\bibnamefont{Monfardini}},
  \bibinfo{author}{\bibfnamefont{N.}~\bibnamefont{Ponthieu}},
  \bibinfo{author}{\bibfnamefont{E.}~\bibnamefont{Barria}},
  \bibinfo{author}{\bibfnamefont{G.}~\bibnamefont{Bres}},
  \bibinfo{author}{\bibfnamefont{M.}~\bibnamefont{Grollier}},
  \bibinfo{author}{\bibfnamefont{G.}~\bibnamefont{Garde}},
  \bibnamefont{et~al.}, \bibinfo{journal}{J. Low Temp. Phys.}
  \textbf{\bibinfo{volume}{184}}, \bibinfo{pages}{816} (\bibinfo{year}{2016}).

\bibitem[{\citenamefont{Mazin et~al.}(2013)\citenamefont{Mazin, Meeker,
  Strader, Szypryt, Marsden, Eyken, Duggan, Walter, Ulbricht, Johnson
  et~al.}}]{Mazin2013}
\bibinfo{author}{\bibfnamefont{B.~A.} \bibnamefont{Mazin}},
  \bibinfo{author}{\bibfnamefont{S.~R.} \bibnamefont{Meeker}},
  \bibinfo{author}{\bibfnamefont{M.~J.} \bibnamefont{Strader}},
  \bibinfo{author}{\bibfnamefont{P.}~\bibnamefont{Szypryt}},
  \bibinfo{author}{\bibfnamefont{D.}~\bibnamefont{Marsden}},
  \bibinfo{author}{\bibfnamefont{J.~C.~v.} \bibnamefont{Eyken}},
  \bibinfo{author}{\bibfnamefont{G.~E.} \bibnamefont{Duggan}},
  \bibinfo{author}{\bibfnamefont{A.~B.} \bibnamefont{Walter}},
  \bibinfo{author}{\bibfnamefont{G.}~\bibnamefont{Ulbricht}},
  \bibinfo{author}{\bibfnamefont{M.}~\bibnamefont{Johnson}},
  \bibnamefont{et~al.}, \bibinfo{journal}{Publ. Astron. Soc. Pac.}
  \textbf{\bibinfo{volume}{125}}, \bibinfo{pages}{pp. 1348}
  (\bibinfo{year}{2013}).

\bibitem[{\citenamefont{Meeker et~al.}(2015)\citenamefont{Meeker, Mazin,
  Jensen-Clem, Walter, Szypryt, Strader, and Bockstiegel}}]{Meeker2015}
\bibinfo{author}{\bibfnamefont{S.~R.} \bibnamefont{Meeker}},
  \bibinfo{author}{\bibfnamefont{B.~A.} \bibnamefont{Mazin}},
  \bibinfo{author}{\bibfnamefont{R.}~\bibnamefont{Jensen-Clem}},
  \bibinfo{author}{\bibfnamefont{A.~B.} \bibnamefont{Walter}},
  \bibinfo{author}{\bibfnamefont{P.}~\bibnamefont{Szypryt}},
  \bibinfo{author}{\bibfnamefont{M.~J.} \bibnamefont{Strader}},
  \bibnamefont{and}
  \bibinfo{author}{\bibfnamefont{C.}~\bibnamefont{Bockstiegel}},
  \bibinfo{journal}{Adaptive Optics for Extremely Large Telescopes 4 -
  Conference Proceedings}  (\bibinfo{year}{2015}).

\bibitem[{\citenamefont{Leduc et~al.}(2010)\citenamefont{Leduc, Bumble, Day,
  Eom, Gao, Golwala, Mazin, McHugh, Merrill, Moore et~al.}}]{Leduc2010}
\bibinfo{author}{\bibfnamefont{H.~G.} \bibnamefont{Leduc}},
  \bibinfo{author}{\bibfnamefont{B.}~\bibnamefont{Bumble}},
  \bibinfo{author}{\bibfnamefont{P.~K.} \bibnamefont{Day}},
  \bibinfo{author}{\bibfnamefont{B.~H.} \bibnamefont{Eom}},
  \bibinfo{author}{\bibfnamefont{J.}~\bibnamefont{Gao}},
  \bibinfo{author}{\bibfnamefont{S.}~\bibnamefont{Golwala}},
  \bibinfo{author}{\bibfnamefont{B.~A.} \bibnamefont{Mazin}},
  \bibinfo{author}{\bibfnamefont{S.}~\bibnamefont{McHugh}},
  \bibinfo{author}{\bibfnamefont{A.}~\bibnamefont{Merrill}},
  \bibinfo{author}{\bibfnamefont{D.~C.} \bibnamefont{Moore}},
  \bibnamefont{et~al.}, \bibinfo{journal}{Appl. Phys. Lett.}
  \textbf{\bibinfo{volume}{97}}, \bibinfo{eid}{102509} (\bibinfo{year}{2010}).

\bibitem[{\citenamefont{{Vissers} et~al.}(2013)\citenamefont{{Vissers}, {Gao},
  {Sandberg}, {Duff}, {Wisbey}, {Irwin}, and {Pappas}}}]{Vissers2013}
\bibinfo{author}{\bibfnamefont{M.~R.} \bibnamefont{{Vissers}}},
  \bibinfo{author}{\bibfnamefont{J.}~\bibnamefont{{Gao}}},
  \bibinfo{author}{\bibfnamefont{M.}~\bibnamefont{{Sandberg}}},
  \bibinfo{author}{\bibfnamefont{S.~M.} \bibnamefont{{Duff}}},
  \bibinfo{author}{\bibfnamefont{D.~S.} \bibnamefont{{Wisbey}}},
  \bibinfo{author}{\bibfnamefont{K.~D.} \bibnamefont{{Irwin}}},
  \bibnamefont{and} \bibinfo{author}{\bibfnamefont{D.~P.}
  \bibnamefont{{Pappas}}}, \bibinfo{journal}{Appl. Phys. Lett.}
  \textbf{\bibinfo{volume}{102}}, \bibinfo{eid}{232603} (\bibinfo{year}{2013}).

\bibitem[{\citenamefont{Goupy et~al.}(2016)\citenamefont{Goupy, Adane, Benoit,
  Bourrion, Calvo, Catalano, Coiffard, Hoarau, Leclercq, Le~Sueur
  et~al.}}]{Goupy2016}
\bibinfo{author}{\bibfnamefont{J.}~\bibnamefont{Goupy}},
  \bibinfo{author}{\bibfnamefont{A.}~\bibnamefont{Adane}},
  \bibinfo{author}{\bibfnamefont{A.}~\bibnamefont{Benoit}},
  \bibinfo{author}{\bibfnamefont{O.}~\bibnamefont{Bourrion}},
  \bibinfo{author}{\bibfnamefont{M.}~\bibnamefont{Calvo}},
  \bibinfo{author}{\bibfnamefont{A.}~\bibnamefont{Catalano}},
  \bibinfo{author}{\bibfnamefont{G.}~\bibnamefont{Coiffard}},
  \bibinfo{author}{\bibfnamefont{C.}~\bibnamefont{Hoarau}},
  \bibinfo{author}{\bibfnamefont{S.}~\bibnamefont{Leclercq}},
  \bibinfo{author}{\bibfnamefont{H.}~\bibnamefont{Le~Sueur}},
  \bibnamefont{et~al.}, \bibinfo{journal}{J. Low Temp. Phys.}
  \textbf{\bibinfo{volume}{184}}, \bibinfo{pages}{661} (\bibinfo{year}{2016}).

\bibitem[{\citenamefont{Bentmann et~al.}(2008)\citenamefont{Bentmann, Demkov,
  Gregory, and Zollner}}]{Bentmann2008}
\bibinfo{author}{\bibfnamefont{H.}~\bibnamefont{Bentmann}},
  \bibinfo{author}{\bibfnamefont{A.~A.} \bibnamefont{Demkov}},
  \bibinfo{author}{\bibfnamefont{R.}~\bibnamefont{Gregory}}, \bibnamefont{and}
  \bibinfo{author}{\bibfnamefont{S.}~\bibnamefont{Zollner}},
  \bibinfo{journal}{Phys. Rev. B} \textbf{\bibinfo{volume}{78}},
  \bibinfo{pages}{205302} (\bibinfo{year}{2008}).

\bibitem[{\citenamefont{{Oto} et~al.}(1994)\citenamefont{{Oto}, {Takaoka},
  {Murase}, and {Ishida}}}]{Oto1994}
\bibinfo{author}{\bibfnamefont{K.}~\bibnamefont{{Oto}}},
  \bibinfo{author}{\bibfnamefont{S.}~\bibnamefont{{Takaoka}}},
  \bibinfo{author}{\bibfnamefont{K.}~\bibnamefont{{Murase}}}, \bibnamefont{and}
  \bibinfo{author}{\bibfnamefont{S.}~\bibnamefont{{Ishida}}},
  \bibinfo{journal}{J. Appl. Phys.} \textbf{\bibinfo{volume}{76}},
  \bibinfo{pages}{5339} (\bibinfo{year}{1994}).

\bibitem[{\citenamefont{Szypryt et~al.}(2015)\citenamefont{Szypryt, Mazin,
  Bumble, Leduc, and Baker}}]{Szypryt2015}
\bibinfo{author}{\bibfnamefont{P.}~\bibnamefont{Szypryt}},
  \bibinfo{author}{\bibfnamefont{B.}~\bibnamefont{Mazin}},
  \bibinfo{author}{\bibfnamefont{B.}~\bibnamefont{Bumble}},
  \bibinfo{author}{\bibfnamefont{H.}~\bibnamefont{Leduc}}, \bibnamefont{and}
  \bibinfo{author}{\bibfnamefont{L.}~\bibnamefont{Baker}},
  \bibinfo{journal}{IEEE Trans. Appl. Supercond.}
  \textbf{\bibinfo{volume}{25}}, \bibinfo{pages}{1} (\bibinfo{year}{2015}).

\bibitem[{\citenamefont{Gao}(2008)}]{Gao2008b}
\bibinfo{author}{\bibfnamefont{J.}~\bibnamefont{Gao}}, Ph.D. thesis,
  \bibinfo{school}{California Institute of Technology} (\bibinfo{year}{2008}).

\bibitem[{\citenamefont{van Eyken et~al.}(2015)\citenamefont{van Eyken,
  Strader, Walter, Meeker, Szypryt, Stoughton, O’Brien, Marsden, Rice, Lin
  et~al.}}]{vanEyken2015}
\bibinfo{author}{\bibfnamefont{J.~C.} \bibnamefont{van Eyken}},
  \bibinfo{author}{\bibfnamefont{M.~J.} \bibnamefont{Strader}},
  \bibinfo{author}{\bibfnamefont{A.~B.} \bibnamefont{Walter}},
  \bibinfo{author}{\bibfnamefont{S.~R.} \bibnamefont{Meeker}},
  \bibinfo{author}{\bibfnamefont{P.}~\bibnamefont{Szypryt}},
  \bibinfo{author}{\bibfnamefont{C.}~\bibnamefont{Stoughton}},
  \bibinfo{author}{\bibfnamefont{K.}~\bibnamefont{O’Brien}},
  \bibinfo{author}{\bibfnamefont{D.}~\bibnamefont{Marsden}},
  \bibinfo{author}{\bibfnamefont{N.~K.} \bibnamefont{Rice}},
  \bibinfo{author}{\bibfnamefont{Y.}~\bibnamefont{Lin}}, \bibnamefont{et~al.},
  \bibinfo{journal}{Astrophys. J. Supp.} \textbf{\bibinfo{volume}{219}},
  \bibinfo{pages}{14} (\bibinfo{year}{2015}).

\bibitem[{\citenamefont{{Gao} et~al.}(2008)\citenamefont{{Gao}, {Daal},
  {Vayonakis}, {Kumar}, {Zmuidzinas}, {Sadoulet}, {Mazin}, {Day}, and
  {Leduc}}}]{Gao2008}
\bibinfo{author}{\bibfnamefont{J.}~\bibnamefont{{Gao}}},
  \bibinfo{author}{\bibfnamefont{M.}~\bibnamefont{{Daal}}},
  \bibinfo{author}{\bibfnamefont{A.}~\bibnamefont{{Vayonakis}}},
  \bibinfo{author}{\bibfnamefont{S.}~\bibnamefont{{Kumar}}},
  \bibinfo{author}{\bibfnamefont{J.}~\bibnamefont{{Zmuidzinas}}},
  \bibinfo{author}{\bibfnamefont{B.}~\bibnamefont{{Sadoulet}}},
  \bibinfo{author}{\bibfnamefont{B.~A.} \bibnamefont{{Mazin}}},
  \bibinfo{author}{\bibfnamefont{P.~K.} \bibnamefont{{Day}}}, \bibnamefont{and}
  \bibinfo{author}{\bibfnamefont{H.~G.} \bibnamefont{{Leduc}}},
  \bibinfo{journal}{Appl. Phys. Lett.} \textbf{\bibinfo{volume}{92}},
  \bibinfo{eid}{152505} (\bibinfo{year}{2008}).

\bibitem[{\citenamefont{Noroozian}(2012)}]{Noroozian2012}
\bibinfo{author}{\bibfnamefont{O.}~\bibnamefont{Noroozian}}, Ph.D. thesis,
  \bibinfo{school}{California Institute of Technology} (\bibinfo{year}{2012}).

\bibitem[{\citenamefont{Golwala}(2000)}]{Golwala2000}
\bibinfo{author}{\bibfnamefont{S.~R.} \bibnamefont{Golwala}}, Ph.D. thesis,
  \bibinfo{school}{University of California, Berkeley} (\bibinfo{year}{2000}).

\bibitem[{\citenamefont{Sac\'ep\'e et~al.}(2008)\citenamefont{Sac\'ep\'e,
  Chapelier, Baturina, Vinokur, Baklanov, and Sanquer}}]{Sacepe2008}
\bibinfo{author}{\bibfnamefont{B.}~\bibnamefont{Sac\'ep\'e}},
  \bibinfo{author}{\bibfnamefont{C.}~\bibnamefont{Chapelier}},
  \bibinfo{author}{\bibfnamefont{T.~I.} \bibnamefont{Baturina}},
  \bibinfo{author}{\bibfnamefont{V.~M.} \bibnamefont{Vinokur}},
  \bibinfo{author}{\bibfnamefont{M.~R.} \bibnamefont{Baklanov}},
  \bibnamefont{and} \bibinfo{author}{\bibfnamefont{M.}~\bibnamefont{Sanquer}},
  \bibinfo{journal}{Phys. Rev. Lett.} \textbf{\bibinfo{volume}{101}},
  \bibinfo{pages}{157006} (\bibinfo{year}{2008}).

\end{thebibliography}

\providecommand{\noopsort}[1]{}\providecommand{\singleletter}[1]{#1}%

\end{document}